\documentclass[10pt,twoside,BCOR7mm,DIV15,headinclude,footexclude,
               cleardoubleempty,idxtotoc]
{scrartcl}

\usepackage{natbib}
\usepackage[font=small,labelfont=bf]{caption}
\usepackage[english]{babel}
\usepackage{graphicx}
\usepackage{hyperref}
\usepackage{scrpage2}
\usepackage{ifthen}
\usepackage{booktabs}
\usepackage{amsmath}
\usepackage{amssymb}
\usepackage{multicol}
\usepackage{float}
\usepackage{hyperref}

\hypersetup{breaklinks=true
,colorlinks=true,linkcolor=black,urlcolor=blue
,citecolor=black}

\addto\captionsenglish{%
}

\makeatletter
\renewcommand{\@biblabel}[1]{}
\renewcommand{\@cite}[2]{%
{#1\ifthenelse{\boolean{@tempswa}}{,#2}{}}}
\makeatother
\ofoot{\thepage}
\ifoot{}

\setcapindent{0em}
\setheadsepline{1pt}

\setkomafont{pagehead}{\normalfont\sffamily}
\setkomafont{pagenumber}{\normalfont\rmfamily}

\makeatletter
\newcommand{\listofcontributions}{\@starttoc{con}}

\newcommand{\l@contribution} {\@dottedtocline{1}{1.5em}{2.3em}}
\makeatother

\newenvironment{contribution}{
\setcounter{section}{0}
\setcounter{figure}{0}
\setcounter{table}{0}
}{
\newpage
\lehead{}
\rohead{}
}

\def\aj{AJ}%
\def\apj{ApJ}%
\def\aap{A\&A}%
\def\mnras{MNRAS}%
\begin{document}

\setlength{\baselineskip}{2.5ex}

\begin{contribution}

\lehead{K.\ R.\ Sokal et al.}

\rohead{The importance of Wolf-Rayet ionization and feedback}

\begin{center}
{\LARGE \bf The importance of Wolf-Rayet ionization and feedback on super star cluster evolution}\\
\medskip

{\it\bf K.\ R.\ Sokal$^1$, K.\ E.\ Johnson$^1$, P.\ Massey$^2$, \& R.\ Indebetouw$^1$}\\

{\it $^1$University of Virginia, United States}\\
{\it $^2$Lowell Observatory, United States}

\begin{abstract}
The feedback from massive stars is important to super star cluster (SSC) evolution and the timescales on which it occurs. SSCs form embedded in thick material, and eventually, the cluster is cleared out and revealed at optical wavelengths -- however, this transition is not well understood. We are investigating this critical SSC evolutionary transition with a multi-wavelength observational campaign. Although previously thought to appear after the cluster has fully removed embedding natal material, we have found that SSCs may host large populations of Wolf-Rayet stars. These evolved stars provide ionization and mechanical feedback that we hypothesize is the tipping point in the combined feedback processes that drive a SSC to emerge. Utilizing optical spectra obtained with the 4m Mayall Telescope at Kitt Peak National Observatory and the 6.5m MMT, we have compiled a sample of embedded SSCs that are likely undergoing this short-lived evolutionary phase and in which we confirm the presence of Wolf-Rayet stars. Early results suggest that WRs may accelerate the cluster emergence. 

\end{abstract}
\end{center}

\begin{multicols}{2}

\section{Introduction}

\begin{figure*}[!t]
\begin{center}
\includegraphics
  [width=\textwidth]{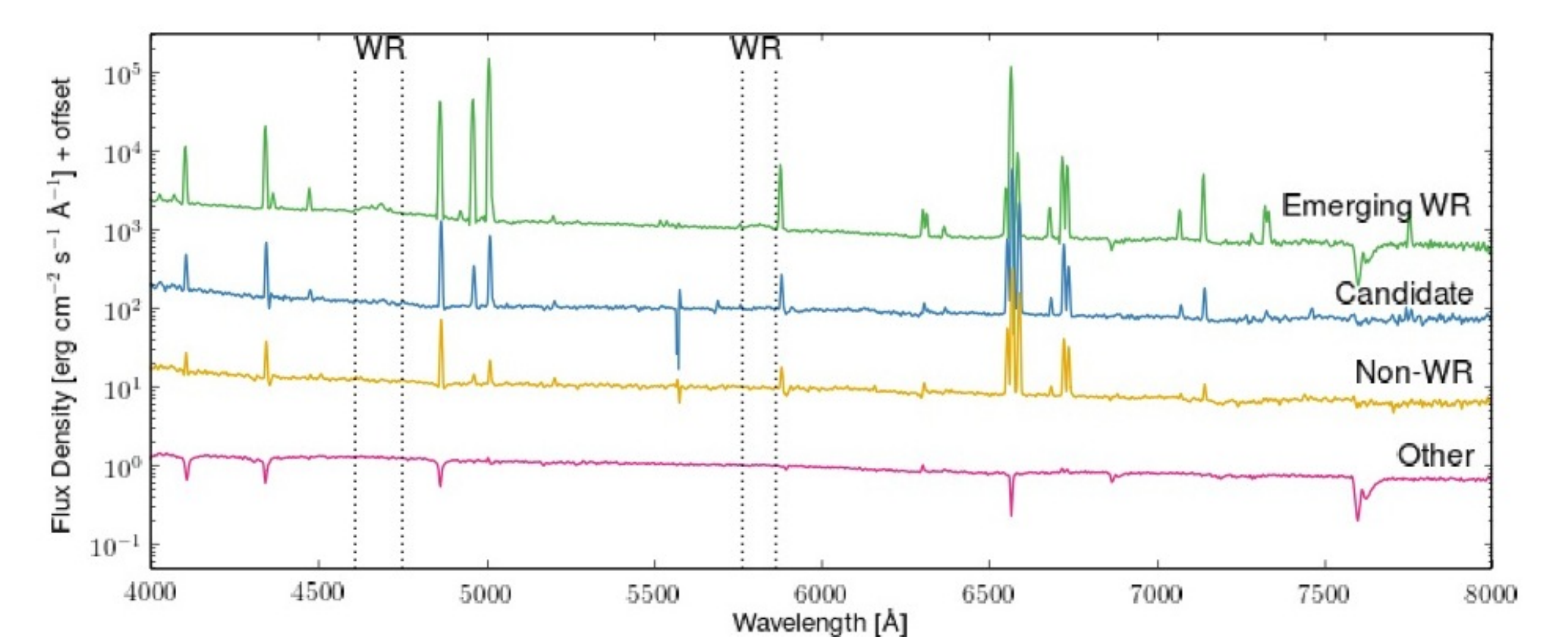}
\caption{An example optical spectrum observed with the 4m Mayall Telescope at KPNO and the 6.5m MMT of each type of class within our sample. Dashed lines show where the broad WR features (`bump') lie.
\label{figure:class-spec}}
\end{center}
\end{figure*}

The highest concentrations of Wolf-Rayet stars (WRs) are found extragalactically in massive and 
super star clusters (SSCs). These bright, blue star clusters have masses as high as 10$^6$ M$_{\odot}$ 
and host hundreds to thousands of massive stars, which interact with each other, in a single dense cluster. These regions are thus equivalent to, or more massive than, the closest well-known example R136 in the Large Magellanic Cloud. As SSCs are so rich,                                                                                                                                                  these are some of the most extreme regions of star formation. 

A cartoon picture of SSC evolution has developed in which SSCs form as scaled-up versions of single massive stars in the Milky Way  \citep{john02}, during which we expect different observable signatures at each stage. A SSC forms from a thick, dense molecular cloud; the proto-SSC is thus embedded in and obscured by an envelope of natal material. Soon, massive stars forming within the cluster ionize this surrounding material. A SSC at this evolutionary stage is detectable as a radio continuum source with thermal emission, which is indicative of this dense young HII region \citep[e.g.][]{kj99}. As the stars continue to evolve and more form, feedback will clear out the surrounding material and will ultimately produce an optically visible cluster.

Yet, how these SSCs emerge from the natal envelope is not yet well understood despite implications for
the fate of the cluster itself as well as for the nearby environment and host galaxy. For instance, a lack of understanding how the ionizing radiation escapes from individual HII regions may be hindering current cosmic simulations \citep{paa11}. The future of a cluster is impacted by the removal of natal gas, as this can effect further star formation efficiencies \citep[even halting further star formation, ][]{bau07} and the cluster's ability to stay bound and thus survive \citep{pfa13}. Understanding cluster emergence has been extremely difficult, largely because
these are messy environments with many physical mechanisms at play. Feedback
processes include: direct stellar radiation; photoionization; pressure from cold, warm (ionized), and hot gas; dust-processed IR radiation; protostellar wind and jets; and later, winds and supernova from massive stars \citep[e.g.][]{lop11}.

Simulations and observations  have recently concentrated on identifying the dominant feedback mechanism.
However, simulations often are limited in some capacity. Modeled star clusters are typically less massive than SSCs, and the input coupling of the feedback to the cloud material is not yet well founded \citep{rp13}.  Fortunately, as simulations are becoming more powerful, they are able pull out more details about how clusters emerge: for instance, in comparing the effects of stellar winds versus photoionization. \citet{dale14} finds that photoionization dominates the energetics during star cluster evolution, yet  the additional inclusion of winds is necessary to get observed morphologies of the produced HII region.

Observationally, a consensus on the dominate feedback mechanism has not been reached. \citet{lop11} compares the pressures due to various feedback mechanisms  (stellar radiation, dust-processed IR radiation, and
the different temperature gas components) and finds that radiation pressure dominates in
30 Dor. Alternatively, an independent study by \citet{pel11} concludes that hot gas dominates instead. Moreover, a larger expanded sample of HII regions in the Magellanic Clouds finds that the warm ionized gas pressure dominates \citep{lop14}.

By highlighting an overlooked yet potential source of feedback instead, we provide a fresh look into this important evolutionary transition through the identification of an emerging massive star cluster. This star cluster is a prime example of this phase as enshrouding natal material is being drastically altered and evacuated by a massive star population containing WRs. 

\section{S26 - discovery of emerging WR clusters}

S26 in NGC 4449 was identified as a partially-embedded radio continuum source \citep{rei08} with a thermal emission component. An extragalactic thermal radio detection of an HII region, which is rather rare \citep{ave11},  indicates youth and either vast size or high density. Archival {\em Hubble Space Telescope} images show that S26 also quite bright optically and currently emerging. When we obtained optical spectra of S26, we discovered a surprising feature know as the WR bump \citep{rei10,sok15} due to integrated stellar emission from WRs. 

Given our previous understanding of timescales, one 
would not expect for WRs to appear until after a star cluster has emerged. Thus their simultaneous presence
with thermal radio emission may suggest that S26 remained embedded until the WRs 
help it emerge \citep{sok15}. Additional evidence for ongoing feedback is also seen in the infrared SED from archival Spitzer and Herschel Space Telescope images
and a possible nebular bipolar outflow in the cluster center \citep{sok15}. Because of S26, we hypothesize that WRs may provide the tipping point in the combined
feedback processes that drive a SSC to emerge \citep{sok15}. 


\section{Finding more emerging WR clusters}


\subsection{Observational survey}


We have carried out an observational survey to identify more clusters like S26. Targeting radio continuum sources with thermal emission similar to S26 in star-forming galaxies, 
we obtained optical spectra with the 6.5m MMT at the Fred Lawrence Whipple Observatory and  the 4m Mayall 
Telescope at Kitt Peak National Observatory to search for the WR bump. In 
line with a classification scheme from \citet{whit14}, we find clusters undergoing the emerging phase via detected radio emission, optical continuum, and optical lines. By searching for emerging clusters with WR features, we are looking specifically for `emerging WR clusters.'

\subsection{Success! and classifications}

Clear detections of the WR bump are found  in many targets -- vastly expanding our sample of emerging WR clusters. However, there are many sources  in which we do not see a clear WR bump, these range from very different objects, to HII regions without any bump whatsoever, and to sources with possible or only nebular WR features. As such, we have classified our sample of radio-selected sources as `emerging WR' if the WR bump is detected, `candidate' if there is a non-significant detection of the WR bump, `non-WR' if the WR bump is  not detected, and `other' if the spectra do not resemble emission line spectra expected of HII regions (Sokal et al. 2015b, in prep).  Example spectra and the distribution of the classes are shown in Fig. \ref{figure:class-spec} and \ref{figure:pie}.
 
 \newpage
 
\begin{figure}[H]
\begin{center}
\includegraphics[width=\columnwidth]{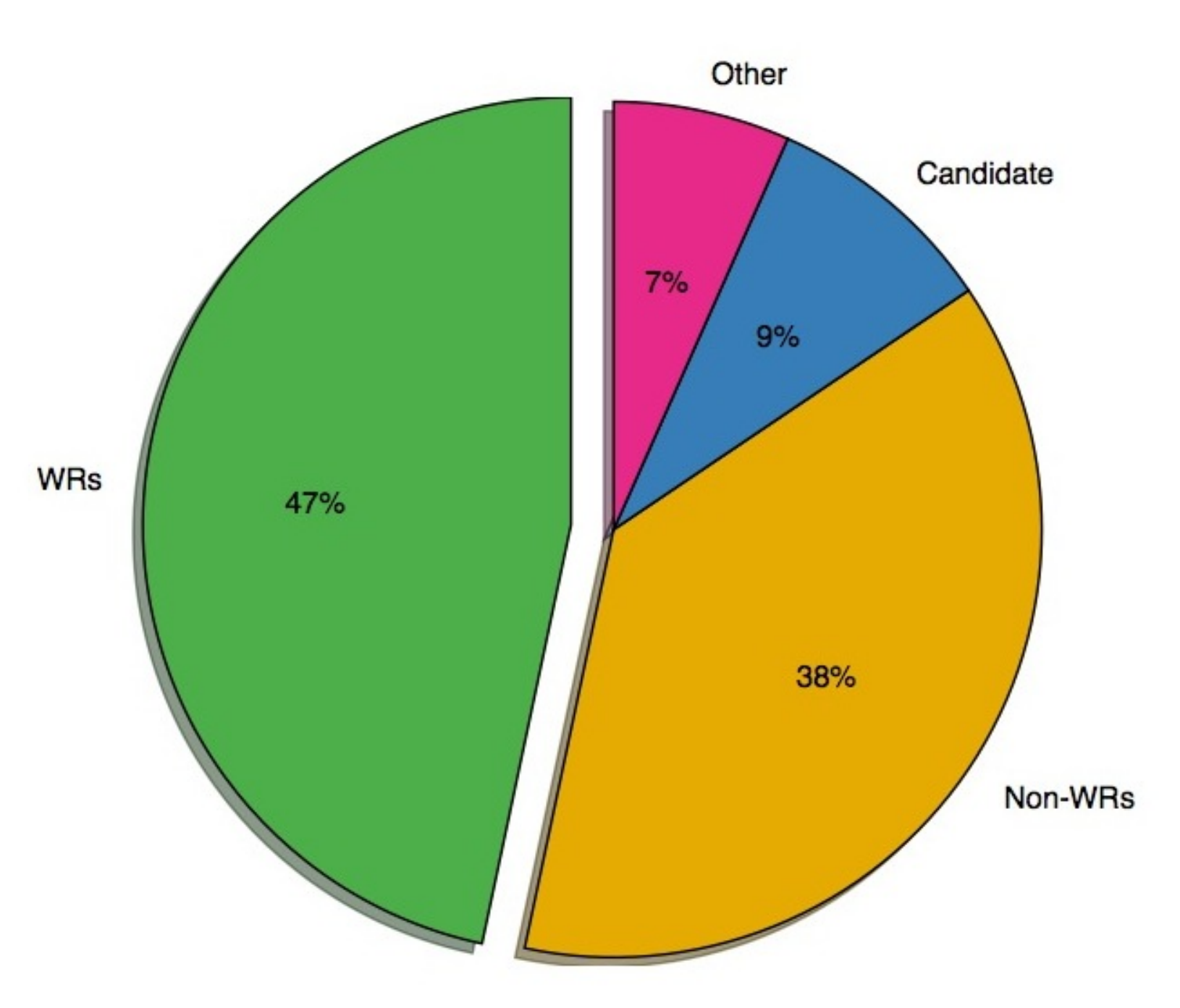}
\caption{The distribution of the classes observed in our sample. Emerging WR clusters are the most common amongst the classes, and if the Other class is omitted, these clusters makes up more that 50\% of the sample. 
\label{figure:pie}}
\end{center}
\end{figure}

\subsection{WRs and cluster evolution}

With this survey, we have identified 21 emerging WR clusters by searching for the 
WR bump in a sample of 45 radio-selected sources. We find we do not preferentially detect or observe either the emerging WR cluster class nor the non-WR cluster class. Classes with and without WR features show similar observed luminosity distributions and span the same parameter space in radio properties. 

Thus, the observed commonality of the WRs is an important result: a clear detection of the WR bump is observed in $\sim$50\% of our radio-selected sample (see Fig. \ref{figure:pie}). We note that just like proving single stars are single, it is difficult to prove that WRs are not  present in a given integrated spectrum -- rather, one can only show that they are. Still, a compelling percentage of the sample hosts WRs. 

Moreover, we have found there may be large differences between these two classes.
The distribution of cluster ages shows that the emerging WR clusters 
 tend to be younger than sources without WR features. Similarly, the emerging WR clusters  in general are found to have lower extinctions, as shown in Figure \ref{figure:extinction}.
Our preliminary results suggest the sources with the highest extinctions do not have WRs (are non-WR clusters) and are also older (Sokal et al. 2015b, in prep). This  may suggest that clusters without significant populations of WRs remain embedded for longer periods of time than clusters which host WR stars. This scenario  is in agreement with the hypothesize derived from S26 by \citet{sok15} that WRs are evolutionally important for a cluster to emerge.


\begin{figure}[H]
\begin{center}
\includegraphics[width=\columnwidth]{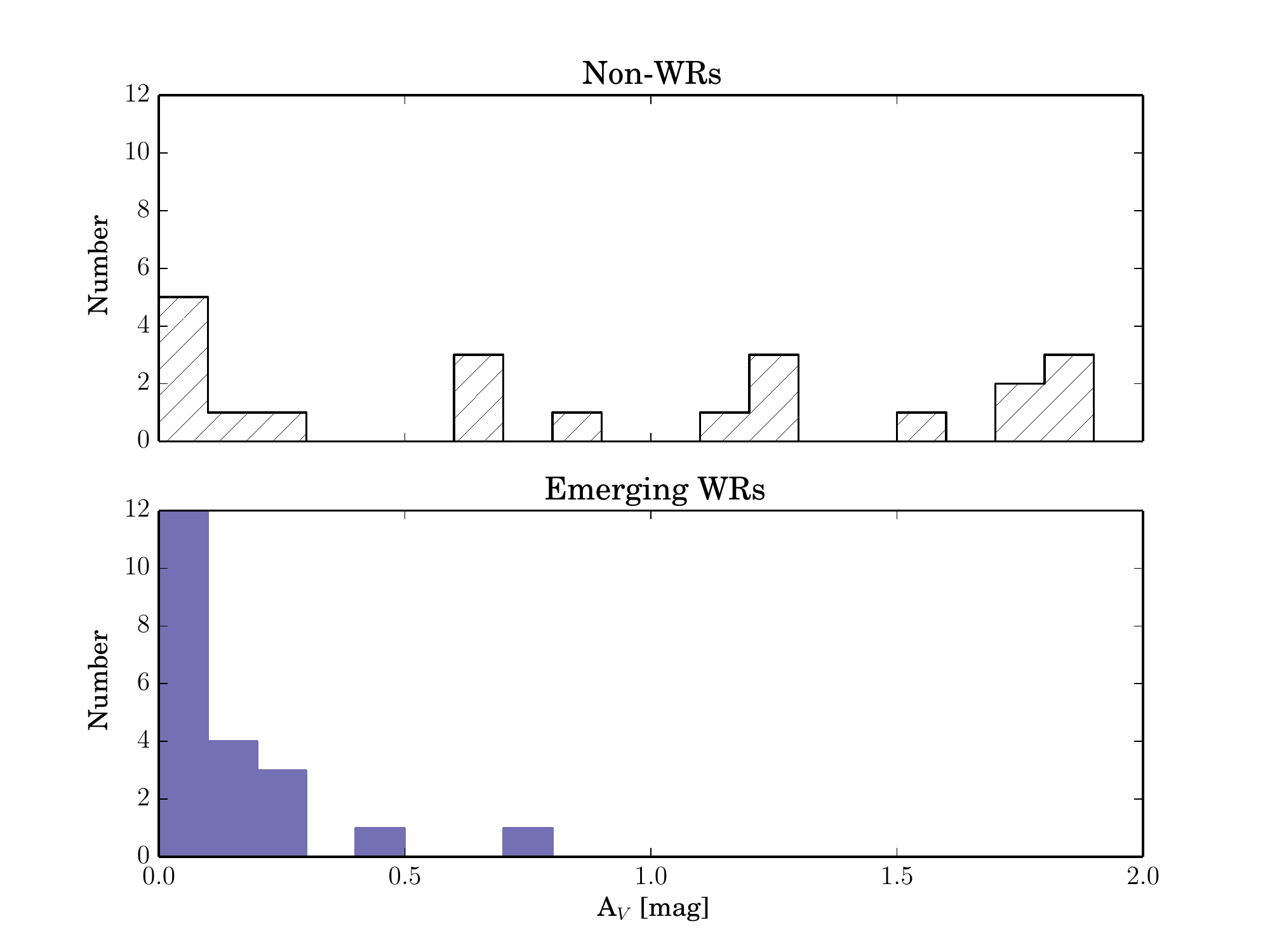}
\caption{The measured extinctions of the emerging WR cluster and non-WR cluster classes. We see that
sources with WR features are less extincted on average.
\label{figure:extinction}}
\end{center}
\end{figure}

K.R.S. is thankful for support provided by the AAS International Travel Grant, Sigma Xi Grants-In-Aid of Research, and observing support from NOAO. K.E.J. acknowledges support provided by the David and Lucile Packard Foundation.

\bibliographystyle{aa} 


\end{multicols}

\end{contribution}


\end{document}